\documentclass[acmsmall,nonacm]{acmart}

\usepackage[]{hyperref}

\usepackage{listings,multicol}
\newcommand{\code}[1]{\lstinline{#1}}
\newcommand{\pcode}[1]{\texttt{#1}}

\usepackage{enumitem}
\setlist{nosep}

\usepackage[english]{babel}
\usepackage{subcaption}
\usepackage{wrapfig}
\usepackage{placeins}

\begin{document}

\title{LOCO: Rethinking Objects for Network Memory}

\author{George Hodgkins}
\affiliation{%
  \institution{University of Colorado, Boulder}
  \city{Boulder}
  \state{Colorado}
  \country{USA}}
\email{george.hodgkins@colorado.edu}

\author{Mark Madler}
\affiliation{%
  \institution{University of Colorado, Boulder}
  \city{Boulder}
  \state{Colorado}
  \country{USA}}
\email{mark.madler@colorado.edu}

\author{Joseph Izraelevitz}
\affiliation{%
  \institution{University of Colorado, Boulder}
  \city{Boulder}
  \state{Colorado}
  \country{USA}}
\email{joseph.izraelevitz@colorado.edu}

\renewcommand{\shortauthors}{Hodgkins et al.}

\begin{abstract}

In this work, we explore an object-based programming model for filling the
gap between shared memory and distributed systems programming. We argue that
the natural representation for resources distributed across a weak memory network (e.g. RDMA
or CXL) is the traditional shared memory object. This concurrent object
(which we call a ``channel'' object) exports traditional methods, but, especially
in an incoherent or uncacheable memory network, stores its state in a
distributed fashion across all participating nodes.  In a sense, the channel
object's state is stored ``across the network''.  

Based on this philosophy, we introduce the Library of Channel Objects (LOCO), a
library for building multi-node objects on RDMA. Channel objects are well-encapsulated
and composable, designed for both the strong locality effects and the weak consistency of
RDMA. Our channel objects have performance
similar to custom RDMA systems (e.g.\ distributed maps), but with a far simpler
programming model amenable to proofs of correctness.

\end{abstract}

\maketitle
\pagestyle{plain}

\section{Introduction}
\label{sec:intro}

The remote direct memory access (RDMA)
protocol is a network protocol which allows a local machine to access the memory of a
remote machine, without communicating with the remote processor (instead, the
memory access is performed by the network card).
Like memory, RDMA exports a load/store
interface, allowing the local machine to copy from or write to remote memory.
Because it largely bypasses the software networking stack on both ends of the
connection, RDMA communication can be quite fast when compared to traditional
network protocols, especially when low-latency communication is required.

Despite its memory-like interface, RDMA is a hardware-accelerated 
networking protocol, and has traditionally been programmed as such --- not as shared memory.  
Most RDMA libraries are a port of a distributed application, encapsulating it
as a single, global, non-reusable instance for accomplishing a given distributed task (e.g.\ consensus~\cite{aguilera-osdi-2020} or distributed storage~\cite{wang-sigmod-2022, dragojevic-nsdi-2014}).  
These systems build their own ad-hoc intermediate layers to manage
RDMA, making composition, especially in the presence of 
RDMA's weak memory consistency guarantees, difficult and error prone.
The libraries (e.g.~\cite{wang-osdi-2020, cai-vldb-2018}) that do
approach an RDMA cluster explicitly as \emph{memory} try to hide the complexities
of a highly non-uniform, weakly consistent network memory in 
a single flat memory space under a distributed shared memory abstraction.
This approach is unlikely to provide optimal performance,
given the community's experience with NUMA-unaware 
applications~\cite{liu-ppopp-2014, tang-hpca-2013, majo-pc-2017}.
Other intermediate layers,
such as MPI~\cite{mpi} or NCCL~\cite{nccl}, 
are abstractions designed explicitly for networks, and forefront
a primarily message passing interface ideal for embarassingly parallel or task-oriented
workflows.  While useful for many applications, such an interface is ill-suited to
irregular and data-dependent workloads, e.g.\ data stores or stateful
transactional systems, for which shared-memory solutions excel~\cite{liu-sigmodrec-2021}.

In this paper, we argue for a new method for programming on networks of
machines connected by RDMA and extensible to other 
weak memory fabrics: fundamentally, a network of RDMA connected machines
can be seen not as a distributed system, but rather as a large, shared memory
machine with extremely weak memory consistency guarantees, strong NUMA effects,
and a relatively reliable transport layer.  This ``scale-up'' perspective,
introduced in~\cite{aguilera-hotos-2019},
is useful for irregular workloads that exhibit data-dependent communication
with unavoidable synchronization, namely, those best suited to shared memory,
but whose working set may not fit in a single node's memory.
From this perspective,
we argue the ideal structure of an RDMA program is as a collection
of highly concurrent, locality-aware objects and data structures tuned for the
heterogeneity and weakness of the memory network.  Given the current trends in
memory technologies and the growth of weak memory fabrics (e.g.\ 
NVidia's NVLink, AMD's Infinity Fabric, or Intel's CXL), 
we expect this ``middle-ground'' environment between shared
memory and networked systems will only grow, and we believe our programming
model to be a strong contender for developing on such systems.

Our implementation of this idea, introduced in this paper, 
is called the \emph{Library of Channel Objects (LOCO)}.  
Each ``channel'' object
is a concurrent object that exports traditional 
methods, but 
stores its state
in a distributed fashion across all participating nodes.  
Familiar examples of supported channels include cross-node
locks, barriers, queues, and maps.
The use of channel objects means both programming
and proof construction
is well encapsulated and highly composable.  For instance, we can use smaller channels, 
such as locks with well-defined semantics, within a larger channel object,
such as a concurrent map. 
Our contributions include:
\begin{itemize}[leftmargin=*]
\item An object library for RDMA, called the Library of Channel Objects (LOCO), which introduces
a channel-based programming model for weak memory networks and provides
support for managing these objects.
\item A collection of core channel objects with a mechanism for allocating, connecting, and composing channels across multiple nodes.
\item A memory consistency mechanism for ensuring relevant 
remote accesses are appropriately synchronized and ordered across channel objects.
\item A provably linearizable key-value store built using our composable primitives.
\item Results demonstrating performance
numbers which compare well with state-of-the-art custom solutions.
\end{itemize}

\section{Background}
\label{sec:background}

This section describes the remote direct memory access protocol (RDMA), and how this
complicated memory network poses programmatic problems addressed by LOCO.

\subsection{Basic RDMA Protocol}

The RDMA protocol directly exposes a machine's memory to remote peers.
The protocol achieves high performance because it offloads most operations to
the network interface card (NIC), traversing only a minimal software stack and
avoiding CPU overheads.

At the application level, RDMA allows a local machine to both \emph{read} from
and \emph{write} to an address on a remote machine. These operations, termed
\emph{one-sided verbs} in RDMA, are managed by the two participating NICs.  

RDMA also
supports supports traditional message-passing with the \emph{send} and
\emph{receive} verbs which require the participation of both CPUs. LOCO supports
these verbs 
but we have not found significant
utility for them in our programming model, and we elide further discussion.

To establish a reliable RDMA connection between two nodes, each participating
node allocates outgoing and incoming request queues at the NIC, called together
a \emph{queue pair} (QP). QPs act as endpoints on the reliable connection. Local
memory which is exposed to the remote node is registered with the NIC as a
\emph{memory region}; its virtual address and protection key must be sent to the
remote node before accesses can be made. 
For reliable
connections, the underlying transport layer ensures verbs are correctly ordered,
missing requests are retransmitted, and timeouts are respected. 
In addition to reliable connections,
RDMA also has support for connectionless and/or unreliable QP links, but they do
not support all one-sided verbs used by LOCO, and so we do not use these link
types.

\subsection{RDMA Memory Consistency}
\label{ssec:consistency}

The base RDMA protocol~\cite{rdma-rfc5040-2007} involves several agents
interacting with memory at once (the local and remote NICs, as well as the local
and remote CPU). The result of concurrent accesses to the same location is not
always clear: RDMA lacks a full, formal, memory consistency, coherence, and
atomicity model, instead specifying the results of racy accesses in terms of
functional effects and allowed reorderings. This lack of cohesion is difficult
for the programmer to reason about, and so LOCO simplifies
within its core channel objects using traditional
mechanisms (e.g.\ atomicity, fences, etc.). To explain these difficulties, we begin by
describing the process of a remote memory write in
RDMA~\cite{rdma-rfc5040-2007}.

An RDMA write is initiated at the local CPU by writing an entry to an
established QP in the NIC's memory over the PCIe bus. This entry includes a pointer to the local
value to be written, or the value itself if
sufficiently small. The local NIC then uses the pointer to access the local
value (or simply copies the value), then sends the write over the network in
a series of packets.  The remote NIC receives the write and, once it verifies
all component packets have \emph{completed} transmission over the network, it
responds to the local NIC with a completion acknowledgement. The remote NIC's
subsequent \emph{placement} of the RDMA write's value in remote memory may
happen during and after completion. 

Given the many units accessing memory, atomicity, coherence, and consistency are
difficult to specify. Instead, the specification gives individual constraints on
the placement of a write (or read), and we leverage them in our library's memory
model. 
First, on machines equipped with Intel DDIO~\cite{intel-ddio-2012} or similar
functionality, the NIC resides within the cache coherence domain, which
dramatically simplifies consistency issues by ensuring NIC operations respect
the TSO global ordering~\cite{intel-mo-2007}.  Next, RDMA packet contents are
not torn by the network, due to the underlying transport protocol. RDMA writes
on the same queue pair are further placed in order.  And finally, a remote write
is fully placed by the NIC before any subsequent remote read or atomic on the
same QP is completed, a feature that can be leveraged to build a fence-type
primitive~\cite{rdma-rfc5040-2007}.

\section{Motivation and Related Work}
\label{sec:motivation}

Prior art using RDMA can generally be categorized into three distinct programming
models, none of which exports reusable primitives suitable for the irregular data
accesses of transactional processing or graph algorithms.

Much prior work in RDMA focuses on \textbf{upper-level primitives}, e.g.\
consensus protocols~\cite{aguilera-osdi-2020,izraelevitz-icpp-2022,jha-tocs-2019,
poke-hpdc-2015, aguilera-podc-2019}, 
distributed maps or 
databases~\cite{dragojevic-sosp-2015,wang-sigmod-2022, dragojevic-nsdi-2014, li-fast-2023,
kalia-sigcomm-2014,barthels-sigmod-2015, alquraan-nsdi-2024, gavrielatos-ppopp-2020},
graph processing~\cite{wang-jpdc-2023}, distributed
learning~\cite{xue-eurosys-2019, ren-hpcc-2017}, 
stand-alone data structures~\cite{brock-icpp-2019,
devarajan-cluster-2020},
disaggregated scheduling~\cite{ruan-hotos-2023,ruan-nsdi-2023}
or 
file systems~\cite{yang-fast-2019,yang-nsdi-2020}.
These works focus on the final application,
rather than considering the
programming model as its own, partitionable problem. As a
result, the intermediate library between RDMA and the exported primitive is
usually ad-hoc and tightly coupled to the application, or effectively
non-existent. In general, these applied, specific, projects manage raw memory
explicitly statically allocated to particular nodes, use ad-hoc atomicity and
consistency mechanisms, and do not consider the possibility of primitive reuse.
This design is not a fundamentally flawed approach, but it does raise the
possibility of a better mechanism, which likely could underlie all the above solutions.

Some works have considered this intermediate layer explicitly,
however, the general approach for this
intermediate layer has been to encapsulate local and remote memory as 
\textbf{distributed shared memory}, that is, a flat,
uniform, coherent, and consistent address space hiding 
the relaxed consistency and
non-uniform performance of the underlying RDMA network. 
These works generally focus on transparently
(or mostly-transparently~\cite{ruan-osdi-2020, zhang-sigmod-2022}) porting
existing shared memory applications.
We argue that this
technique, either with purely
software-based virtualization~\cite{gouk-atc-2022, cai-vldb-2018,
wang-osdi-2020, zhang-sigmod-2022, ruan-osdi-2020}, or by extending
hardware~\cite{calciu-asplos-2021}, 
is unlikely to gain traction because the performance will always be
worse than an approach which takes into account the underlying memory network.

Other programming models have simply used RDMA
to implement existing distributed system abstractions.  For example,
both MPI~\cite{mpi} and NCCL~\cite{nccl} can use RDMA for inter-node communication.  
However, fundamentally, these are \textbf{message passing programming models}
with explicit send and receive primitives.
While MPI does support some remote memory accesses,  
this support is best seen as a zero-copy send/receive
mechanism where synchronization is either coarse-grained and inflexible,
or simply nonexistent. 
While message-passing is well-suited for dataflow applications (e.g.\ machine
learning and signal processing) and highly parallel scale-out workloads
(e.g.\ physical simulation), it is less useful for workloads 
that exhibit data-dependent communication~\cite{liu-sigmodrec-2021}, such as transaction
processing or graph computations. In these applications, cross-node
synchronization is unavoidable and unpredictable, so the ideal performance
strategy shifts from simply avoiding synchronization to minimizing contention,
accelerating synchronization use, and reducing data movement.  

Compared to prior art, 
LOCO aims to build composable, reusable, and performant primitives for 
complicated memory networks, suitable for irregular workloads.
No such option currently exists in the literature.

\section{Design}
\label{sec:design}

Our Library of Channel Objects (LOCO) is functionally an extension
of the normal shared memory programming model, that is, an object-oriented
paradigm, onto the weak memory network of RDMA.
LOCO provides the ability to encapsulate
network memory access within special objects, which we call \emph{channels}. 
Channels are similar to traditional shared memory objects, in that
they export methods, control their own memory and members, and manage
their synchronization. However, unlike traditional shared memory objects,
a single channel may use memory across multiple nodes, including
both network-accessible memory and private local memory. Examples of some
channel types (classes) in LOCO include cross-node mutexes, barriers, queues,
and maps. 

A LOCO application will usually consist of many channels (objects) of many different
channel types (classes). In addition, each channel can itself instantiate member sub-channels
(for instance, a key-value store might include several mutexes as sub-channels
to synchronize access to its contents).  We argue that such a system of channels
makes it significantly easier to develop applications on network memory, without
sacrificing performance.

This section describes the programming model LOCO provides for developing
channels.

\subsection{Channel Overview}
\label{ssec:chan-ov}

First, channels are \textbf{named}: to communicate over a channel, each
participating node constructs a local channel object, or \emph{channel
endpoint}, with the same name. Each channel endpoint allocates zero or more
named local regions of network memory when it is constructed, and metadata
necessary to access these local memory regions is delivered to the other
endpoints during the setup process.

In addition, channels are \textbf{composable}. Each channel contains zero or
more sub-channels, which are namespaced under their parent. This feature is used
to easily compose channel functionality. 

For instance, one channel which we frequently use to build larger channels is
the SST (Shared State Table), first introduced in Derecho~\cite{jha-socc-2017,
jha-tocs-2019} as a non-reusable primitive. The SST is an array of
single-writer, multiple-reader registers, with one
register per participant.  Nodes can write to their
registers and push the contents remotely, or read, locally, the values of others' registers.

\begin{wrapfigure}[42]{r}{0.45\textwidth}
\centering
\begin{subfigure}{0.45\textwidth}
\begin{lstlisting}[basicstyle={\ttfamily\scriptsize}]
class barrier : public loco::channel {
	unsigned count,num_nodes;
	loco::sst_var<unsigned> sst;
	public:
	void waiting() {
		// complete all outstanding RDMA 
		// operations (Section *@{\color{red}\ref{ssec:fence}}@*)
		mgr()::fence(); 
		count++; // increment our counter
		sst.store_mine(count);
		sst.push_broadcast(); //and push
		bool waiting = true; 
		while(waiting){  // wait for others 
			waiting = false; // to match
			for (auto& row : sst) {
				if (row.load() < count){
					waiting = true; 
					break;} 
			}
		}
	}
	barrier(channel* parent, 
	 string name,manager& cm,int num):
	 channel(parent, name, cm, 
	 channel::expect_num(num-1)), *@\label{cd:bar-expect}@*
	 sst(this,"sst",cm)){ 
		count=0; num_nodes=num;
		channel::join(); *@\label{cd:chan-join}@*
	}
};
\end{lstlisting}
\caption{Complete C++ code for the network barrier, a simple channel object.}
\label{lst:barrier}
\end{subfigure}

\begin{subfigure}{0.45\textwidth}
\centering
\begin{lstlisting}[basicstyle={\scriptsize\ttfamily}]
int main(int argc, char** argv) {
	map<uint32_t, string> hosts; 
	int node_id, num_nodes; *@\label{cd:host-map}@* 
	loco::parse_hosts(&hosts,
	 &node_id,&num_nodes,argv[1]);
	vector<timespec> lats;
	loco::manager cm(ip_addrs, node_id); *@\label{cd:mgr-init}@* 
	loco::barrier bar("bar", cm, num_nodes); *@\label{cd:bar-init}@*
	cm.wait_for_ready(); *@\label{cd:mgr-wait}@*
	for(int i=0; i<TEST_ITERS; ++i){
		timespec t0 = clock_now();
		bar.waiting(); *@\label{cd:bar-wait}@*
		timespec t1 = clock_now();
		lats.push_back(t1 - t0);
	}   
	cout<<"Avg latency:"<<
	 accumulate(lats.begin(),
	 lats.end(),0.0))/lats.size();
} *@\label{cd:end}@* 
\end{lstlisting}
\caption{A simple (complete) LOCO application measuring barrier latency.}
\label{lst:bar-lat}
\end{subfigure}
\caption{LOCO barrier code}
\end{wrapfigure}

Figure~\ref{lst:barrier} shows our implementation of a barrier channel, based
on~\cite{gupta-clustr-2002}, using a SST sub-channel. As with a traditional
shared memory barrier, it is used synchronize all participants at a
certain point in execution. For each use of the barrier, participants increment
their local, private, count variable, then broadcast the new value to others
using their register in the SST. They then wait locally to leave the barrier until all
participants have a count in the SST not less than their own.

This example shows how channels aid in encapsulation of complex communication.
The barrier does not allocate network memory or perform remote accesses
directly, but instead relies on the SST sub-channel. This allows the
barrier implementation to be very simple: this code is a
complete implementation of a single-threaded barrier in LOCO.

\subsection{Channel Setup}
\label{ssec:chan-setup}

Figure~\ref{lst:bar-lat} shows a complete example LOCO application: a microbenchmark
which repeatedly waits on the barrier (Line~\ref{cd:bar-wait}) and measures its
latency. At line~\ref{cd:mgr-init}, we construct the \code{manager} object from
a set of (ID, hostname) pairs. The \code{manager} establishes connections with
peers and mediates access to per-node resources: peer connections, a shared
completion queue, and network-accessible memory.

The \code{manager} is then used to construct channel endpoints, in this case the
barrier and its sub-channels (Line~\ref{cd:bar-init}). Note that the barrier has
a name \code{"bar"}, which must match the name of the remote barrier endpoints
to complete the connection.  We use a `/' character to denote a sub-channel
relationship (e.g., the full name of the SST in the barrier object is
\code{"bar/sst"}, with component \code{owned_var}s named \code{"bar/sst/ov0"} etc.), 
and a `.' character to denote a component memory region.

When a channel endpoint is constructed, it initializes its local state 
including subchannels, creates
local memory regions, and indicates by name what memory regions it expects other
participants to provide.
Then, it sends a \emph{join} message
(Line~\ref{cd:chan-join}) to each peer with the channel name and the list of
memory regions it expects that peer to provide.

When a peer receives a join message, it first checks if a channel endpoint with
the same name exists locally, and ignores the message if not (in other words,
peers may not participate in all channels). If it finds a matching
endpoint, it verifies its allocated memory regions match those
requested, and returns
a \emph{connect} message containing metadata necessary to access the requested
regions. 
Channels can also register callbacks which run when join and/or
connect messages are received; these are used to create per-participant
sub-channels or memory regions (see Section \ref{ssec:sst} for an example).

\section{Channels}
\label{sec:channels}

The previous section describes the core functionality LOCO provides for building
channels. Using this functionality, we developed a set of core
channels which provide useful primitives for remote memory access,
synchronization, and message passing. This section describes the design of these
channels.

\subsection{Channels for Memory Access}

\subsubsection{Regions and Variables}

\begin{sloppypar}

The basic building block of most LOCO channels is the \code{shared_region},
which allocates a symmetric region of memory on each participants. Each
participant can read and write all other nodes' regions using addresses
offset into the region. Due to a lack of
consistency guarantees, the \code{shared_region} requires additional
synchronization (e.g.\ locks) or usage constraints to be useful.

One such constraint is encapsulated in the \code{owned_var}, a single-writer multi-reader
register. Each \code{owned_var} has a single authoritative copy stored at the
"owner," and cached copies at all other participants, which can be updated
either by a write from the owner (a "push") or reads from non-owners (a "pull"),
depending on the requirements of higher-level channels. The most performant
strategy for guaranteeing atomicity depends on the size of the contained value.
For values of the CPU architecture's atomic word size or smaller, we only need
to ensure that they are aligned to their size, and accesses will be inherently
atomic. For values larger than the atomic word size, we attach a checksum of the
value, and readers retry on a mismatch.

In addition to the single-writer \code{owned_var}, we provide a multiple-writer,
multiple-reader \code{atomic_var} channel which is restricted to values of the
CPU-atomic word size. The \code{atomic_var} has a single ``official'' copy
hosted at one participant, and cached copies at all other participants. The
primary purpose of this channel is to expose atomic operations on remote memory,
but it also supports overlapping reads and writes from any participant.

\end{sloppypar}

\subsubsection{The SST}
\label{ssec:sst}

\begin{wrapfigure}[11]{r}{0.45\textwidth}
  \begin{center}
    \includegraphics[width=0.45\textwidth]{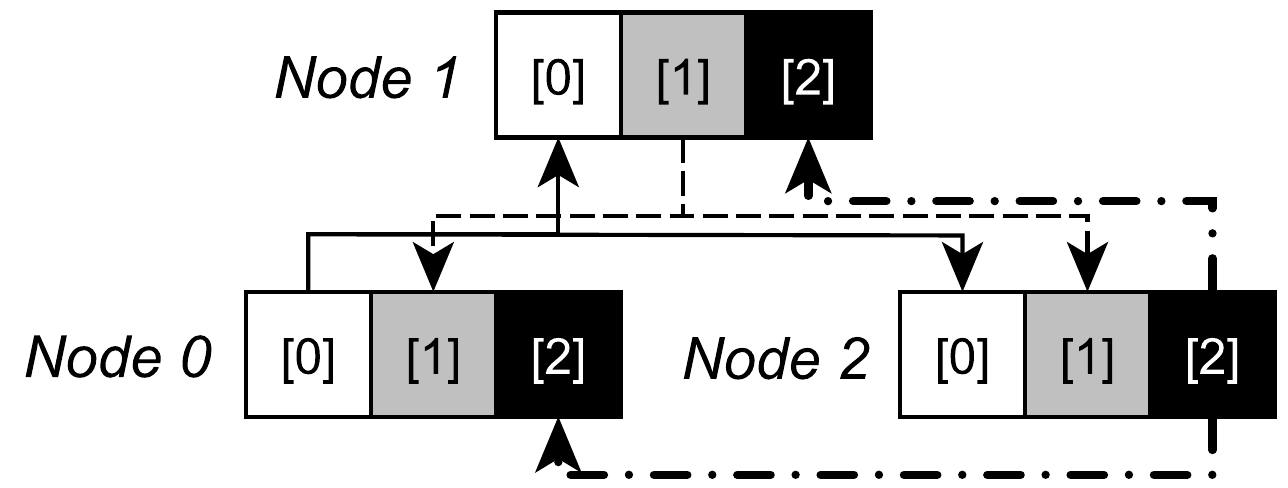}
  \end{center}
\caption{An SST with three participants. Arrows represent \code{owned_vars},
pointing from the writer to readers.}
\label{fig:sst-diag}
\end{wrapfigure}

The SST channel described in the previous section is built out of
\code{owned_var}s. Each participant is a writer on one \code{owned_var} and a
reader on all others, as shown in Figure~\ref{fig:sst-diag}. The design of the
SST showcases the advantages of composable channels: a local SST endpoint simply
consists of a map from node IDs to \code{owned_var} endpoints. Since there is
one endpoint per participant, and the number of participants is not known when
the SST is constructed, new endpoints are constructed and added to the map by a
callback that runs whenever a peer joins the channel.
 
\subsection{Asynchronous Operations}
\label{ssec:async}

To support asynchronous operations on channels, LOCO provides
an \code{ack\_key} object, which is used to query on operation completion.   
When an asynchronous operation, for instance,
an SST broadcast from one writer to all other participants,
is initiated, the caller receives an \code{ack\_key} object
as a return value.  To determine if the broadcast is completed,
the caller can use the \code{query} method to determine if the broadcast
is completed.  Individual \code{ack\_key}'s can be unioned together,
allowing a higher level operation (e.g.\ SST broadcast)
to build its \code{ack\_key} from 
its component operations (e.g.\ individual remote writes on different variables).
Internally, \code{ack_key}'s consist of a set of indexed RDMA operations,
and \code{query}'s check LOCO's internal tracking mechanism to see if the operation
has been completed.
In addition to providing support for asynchronous operations,
\code{ack\_key} infrastructure is also used to implement LOCO's memory
consistency model.

\subsection{Memory Consistency}
\label{ssec:fence}
As described in Section~\ref{ssec:consistency}, RDMA's specification of memory
ordering is weak and unintuitive. To tame this complexity with standard
synchronization primitives, we need a mechanism to guarantee that one node's changes from before
the synchronization are visible after, or, put another way, we need to
induce a synchronizes-with edge in the cross-node happens-before
order. In LOCO, this edge is built using one-sided \emph{fence} primitives
which generalize on the declarative semantics of an existing formal
specification of RDMA operations on our target hardware~\cite{ambal-oopsla-2024}.

A LOCO fence ensures that remote operations
from before the fence will be placed and visible before any subsequent operations.
However, the scope of the ``remote operations'' and ``subsequent operations'' can vary
for performance considerations.
\begin{description}[leftmargin=*]
\item[Pair-only Fence.] Ensures that prior operations are complete before
subsequent ones as long as both operations originate from the calling thread and are placed
on the same remote peer.
\item[Thread-only Fence.] Ensures that prior operations are complete before
subsequent ones as long as both operations originate from the calling thread, regardless
of target peer.
\item[Global Fence.] Ensures that prior operations are complete before
subsequent ones as long as both operations originate from the calling node (regardless of thread or peer).
\end{description}
Cross-node 
synchronize-with edges are induced from a release-write (i.e.\ a write followed by a fence)
to a relaxed local or remote read (i.e.\ no acquire fences are required).  
Depending on the outstanding operations from the local node/thread and the fence scope, 
the fence can range from simply waiting on an \code{ack_key} (pair-only fence)
up to a zero-length remote read from all local threads to all remote nodes with unfenced operations (global fence).
As LOCO tracks outstanding operations, it dynamically chooses the best performing implementation
to accomplish the fence.  

\subsection{Complex Channels}
\label{ssec:more-channels}

Having described the base LOCO channels and its memory consistency
model, our higher level primitives can be straight-forwardly described
using object-oriented language.
\begin{sloppypar}
\begin{description}[leftmargin=*]
\item[Ticket lock.] An implementation of the classic ticket lock algorithm~\cite{mellor-crummey-acmtrans-1991} over network memory.
Nodes use a remote atomic fetch-and-add on a \code{next\_ticket}
value to acquire a ``ticket'', then blocks until a separate \code{now\_serving}
value becomes equal to that ticket (\code{now\_serving} is incremented each time
the lock is released).  Both \code{next\_ticket} and \code{now\_serving} are \code{atomic\_var}s.
The mutex also provides mutual exclusion between local threads, as well as a
mechanism for fast local handover between threads when the lock has already been
acquired on the local node.  LOCO fences used on release and specified by caller.
\item[Barrier.] The channel described in Section~\ref{ssec:chan-ov}, with a global
fence for synchronization.
\item[Ringbuffer.] An asynchronous message-passing primitive that efficiently
implements a one-to-many broadcast.  The ringbuffer is an array of
\code{owned_var}s
with a custom atomicity mechanism to allow for mixed-size messages, and receiver
acknowledgements are sent via SST to allow buffer reuse.  Similar to the buffer
used in FaRM~\cite{dragojevic-nsdi-2014}.
\item[Shared Queue.] A globally-consistent FIFO queue: all participating nodes can both push
to and pop from the queue, with each pop corresponding to exactly one push. The
channel consists of \code{atomic\_var} instances representing head and tail
indices, and memory regions representing the entries of the the queue, striped
across participants.  An adaptation of the shared memory cyclic ring queue~\cite{morrison-ppopp-2013}
for RDMA.
\end{description}
\end{sloppypar}

\section{Example Application}
\label{sec:applications}

In this section, we describe an example LOCO application: a key-value
store. An additional application,
a model of a hardware control loop for distributed control,
can be found in Appendix~\ref{app:powcon}.

\begin{wrapfigure}[16]{r}{0.45\textwidth}
  \begin{center}
    \includegraphics[width=0.44\textwidth]{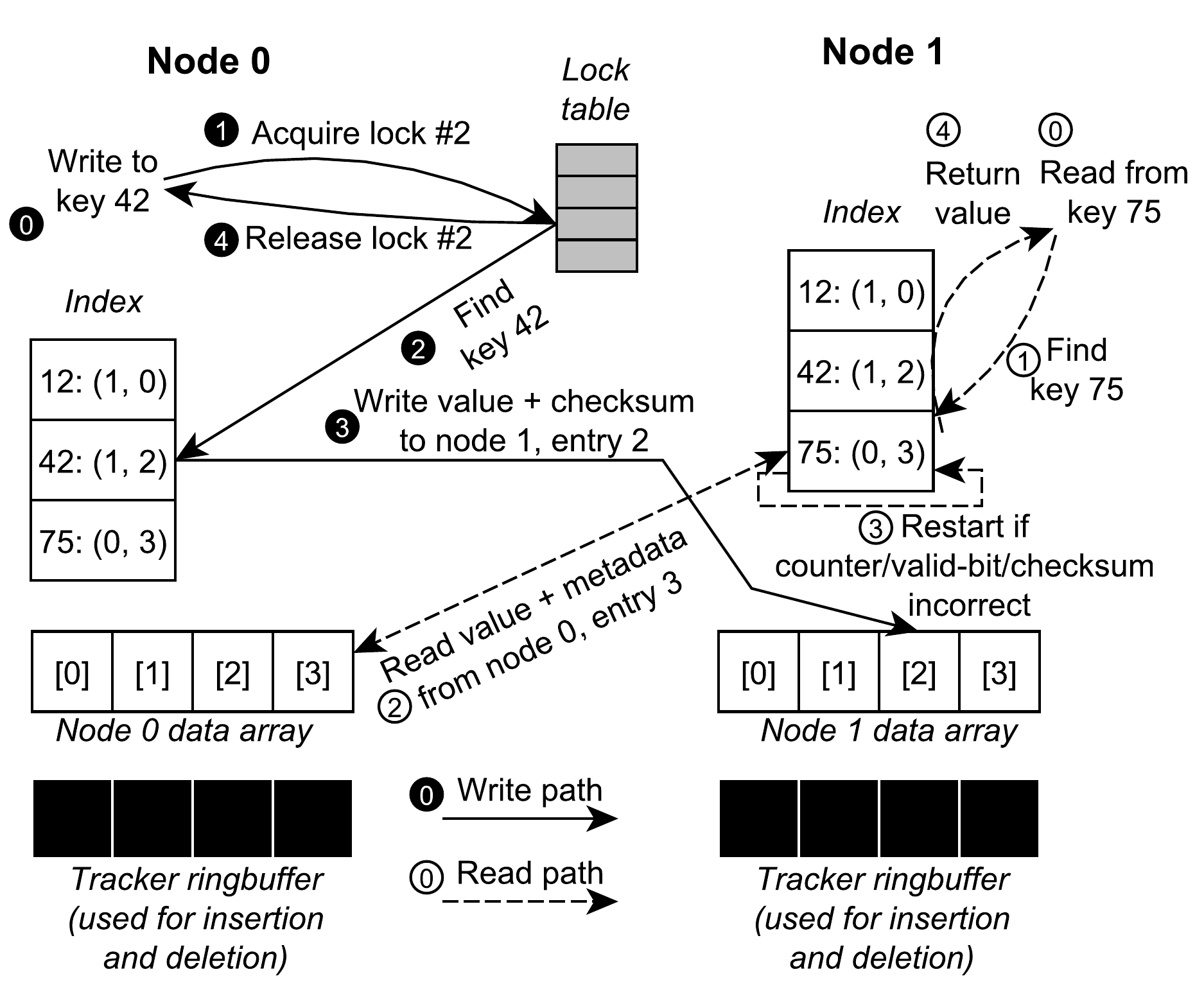}
  \end{center}
 \caption{Read and write operations in the \code{kvstore}.}
 \label{fig:kvstore-diag}
\end{wrapfigure}

Our \pcode{kvstore} channel is a
distributed key-value store with a lookup operation that takes no locks, and
insertion, deletion, and update operations protected by locks. 
 Lookup and update
are depicted in Figure \ref{fig:kvstore-diag}. 
Each node
allocates a remotely-accessible \code{shared_region}
which is used to
store values and consistency metadata (a checksum for atomicity, a counter
for garbage collection, and a valid bit).
Each node also maintains a local index (a C++ \code{unordered_map}), protected
by a local reader-writer lock, which records the locations of all keys in the
\code{kvstore} as \code{(node_id, array_index)} pairs, along with a counter
matching the one stored with the data. The
\pcode{kvstore} is provably linearizable, with an informal proof given in
Appendix~\ref{sec:proof} --- our proof is simplified by leveraging the mutual
exclusion and compositional properties of LOCO objects.  Almost all RDMA
maps~\cite{wang-sigmod-2022, li-fast-2023,
kalia-sigcomm-2014,barthels-sigmod-2015,lu-taco-2024} lack any formal safety
specification (we are only aware of three~\cite{dragojevic-nsdi-2014},\cite{alquraan-nsdi-2024},\cite{gavrielatos-ppopp-2020}), 
likely due to difficulties in encapsulation which the LOCO philosophy solves.

Updates to the indices are protected by an array of \pcode{ticket\_lock}s. 
When a node tries to insert or delete a
key, it first acquires the lock with index \pcode{key \% NUM\_LOCKS}. It
then looks the key up in its local index. In the case of an insertion, if the
key does not yet exist, the node first writes the value to a free slot in its
local data array with the valid bit unset, increments the counter corresponding
to that slot, updates the checksum,
and then broadcasts the value's location and counter to other
nodes on a \code{ringbuffer} called the \emph{tracker}. Each node monitors the
set of other nodes' trackers with a dedicated thread, which applies requested
updates to the local index and then acknowledges the message.  The inserter
waits until all nodes have acknowledged its message, meaning the location of
the key is present in all indices, and then marks the entry valid and releases
the lock.  Deletion is the reverse under the lock; marking the entry invalid, then broadcasting
the deletion, and removing the entry once all nodes have acknowledged it. 

To update the value mapped to a key, a node takes the lock corresponding to that
key and looks up its location in the local index.  If it exists, it writes the
new value to that location (retaining the counter and valid bit), 
updates the checksum, then releases the lock. This write is fenced (see
Section~\ref{ssec:fence}), to ensure it is ordered with the
subsequent lock release. 

To retrieve the value mapped to a key, a node need not take a lock, but simply
looks up the key in the local index, failing if it is not found, and reads the
value and accompanying metadata from their location on the corresponding node.
If the checksum is incorrect due to a torn update, it retries.
If the valid bit is not set (indicating an incomplete insertion/deletion),
or the counter mismatches (indicating a stale local index),
the reader can safely return \pcode{EMPTY}.

\section{Performance Evaluation}
\label{sec:eval}

In this section, we evaluate the performance of two LOCO applications: a
transactional locking benchmark and a key-value store.

All results were collected using \code{c6525-25g} nodes on the Cloudlab
platform~\cite{cloudlab-machines}. These machines each have a 16-core AMD 7302P
CPU, running Ubuntu 22.04. Nodes communicate over a 25 Gbps Ethernet
fabric using Mellanox ConnectX-5 NICs. 

\subsection{Transactional Locking}

\begin{wrapfigure}[16]{r}{0.45\textwidth}
\begin{center}
\begin{subfigure}{.43\textwidth}
\includegraphics[width=\textwidth]{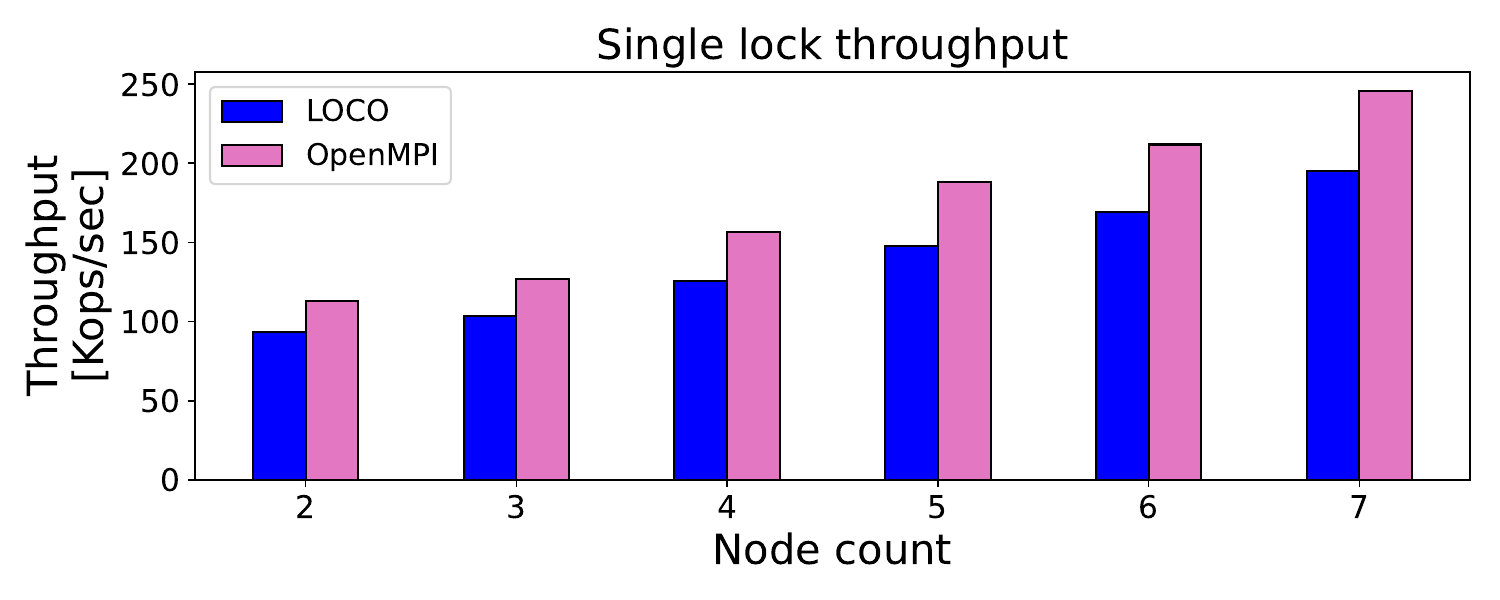}
\end{subfigure}
\begin{subfigure}{.43\textwidth}
\includegraphics[width=\textwidth]{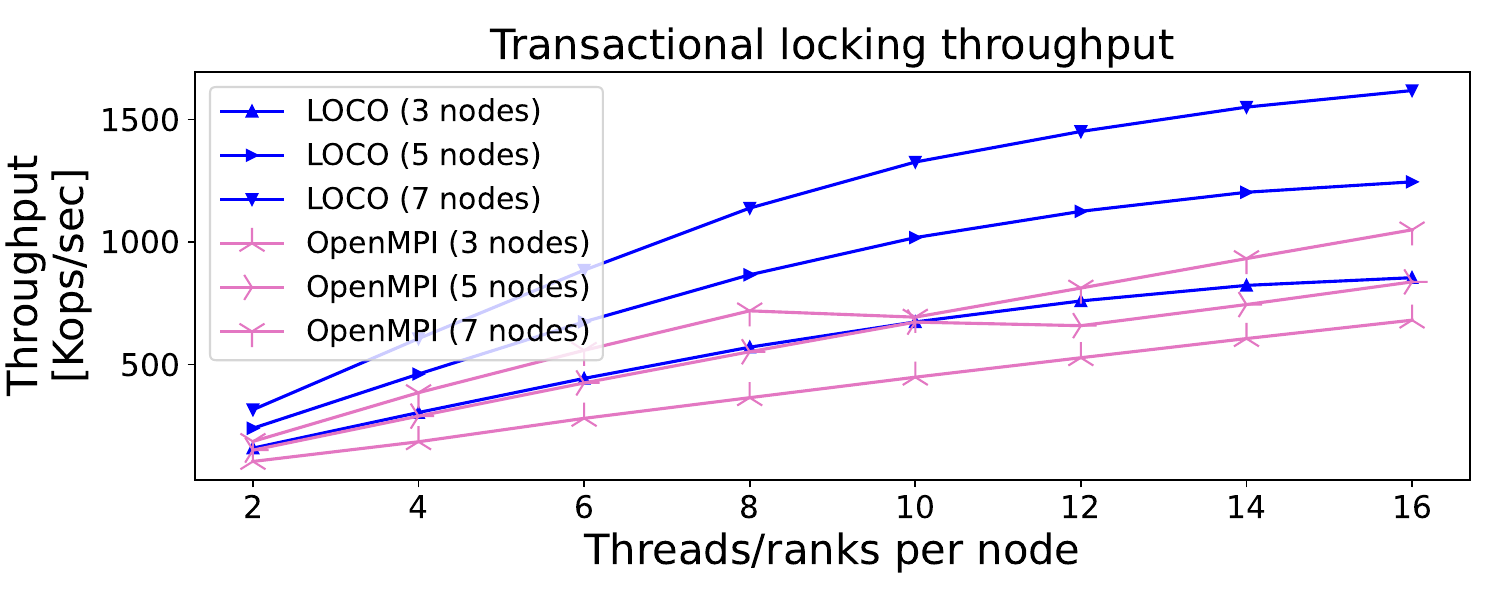}
\end{subfigure}
\caption{Throughput of single-lock and transactions
in OpenMPI and LOCO.}
\label{fig:lock-bench}
\end{center}
\end{wrapfigure}

In this section, we compare the performance of LOCO to the RDMA APIs provided by
OpenMPI~\cite{openmpi} on tasks involving contended synchronization. We compare
against OpenMPI version 5.0.5, using RoCE support provided by the PML/UCX
backend. Results for both benchmarks are shown in Figure~\ref{fig:lock-bench}
(geomean of five 20-second runs). 

First, we measured the throughput of a contended single-lock critical section
(lock-protected read-modify-write) at different node counts, with one
rank/thread per node. Here, OpenMPI has a consistent advantage,
likely due to extensive optimization and a more managed environment. 

Then, we measured the throughput of a transactional critical section, which
acquires the locks corresponding to two different accounts (array entries), and
transfers a randomly generated amount between them. We use 100 million accounts.
For intra-node scaling, LOCO creates multiple threads, while OpenMPI creates
separate ranks (MPI processes), due to MPI's limited support for multithreading
within a rank.

For LOCO, we create an array of \code{atomic_vars} holding account values, striped
across participants. For OpenMPI, we distribute the accounts across 341 windows
(symmetrically allocated regions of remote memory, each associated with a
single lock per rank); 341 is the maximum supported.
To ensure a fair comparison, LOCO uses at most 341 locks per
thread. 

LOCO outperforms OpenMPI on transactional locking, despite the fact
that we use an equal number of locks and their lock performs better in
isolation. We believe this is due to the tight coupling between memory windows
and locks in MPI: 
windows likely have a one-to-one correspondence with RDMA
memory regions in the backend, and performing operations on many small memory
regions is slower than large ones due to NIC caching
structures~\cite{kong-nsdi-2023}.
LOCO avoids this penalty by disassociating regions and locks in its object system, while
also merging regions into 1 GB huge pages in the backend. 

\subsection{Key-Value Store}

\begin{figure*}
\centering
\begin{subfigure}{.5\textwidth}
\includegraphics[width=\textwidth]{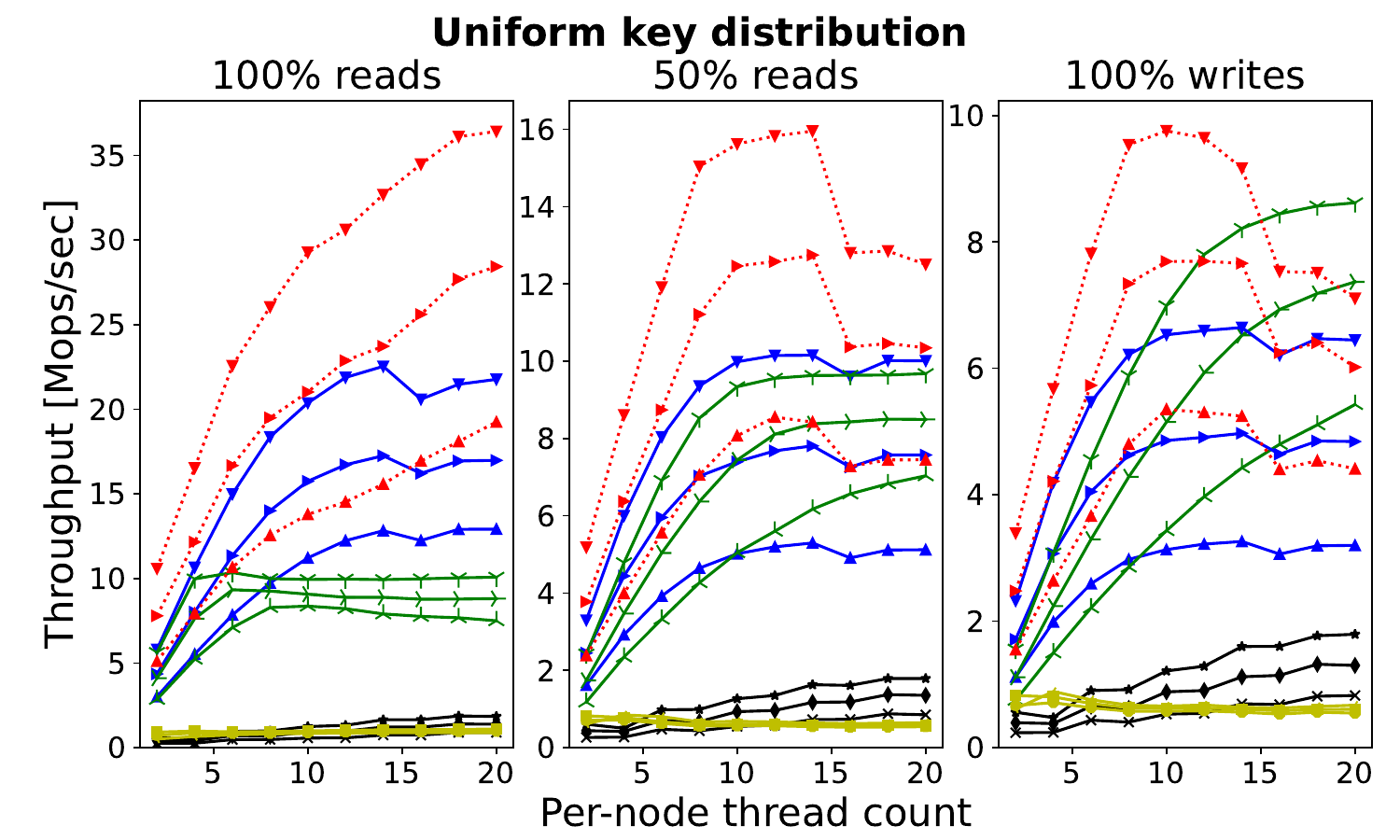}
\end{subfigure}%
\begin{subfigure}{.5\textwidth}
\includegraphics[width=\textwidth]{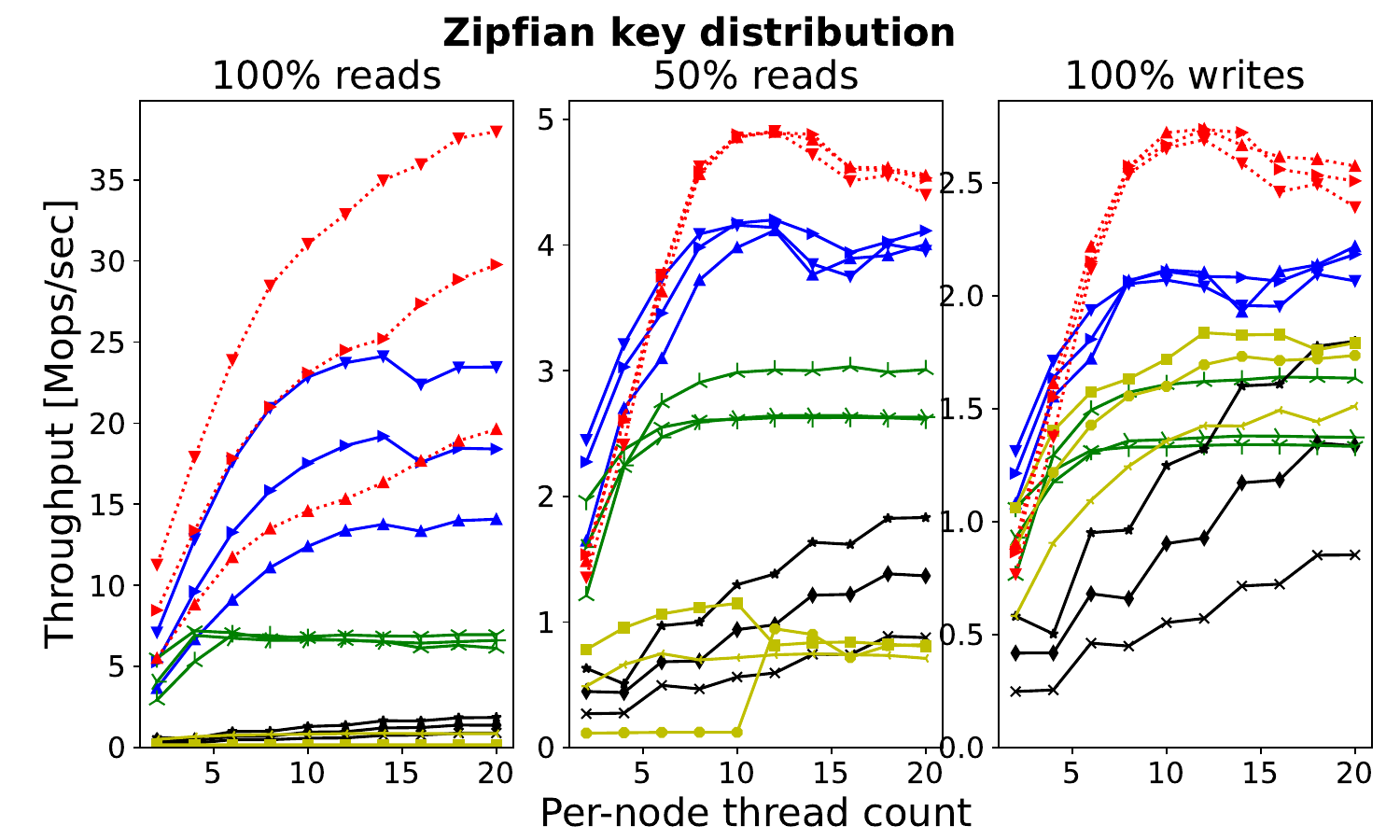}
\end{subfigure}
\includegraphics[width=.8\textwidth]{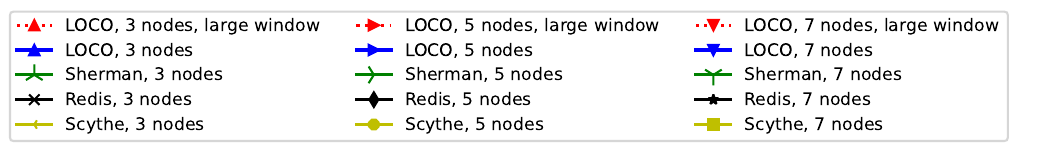}
\caption{Throughput comparison of RDMA key-value stores.}
\label{fig:kvstore-perf}
\end{figure*}

We further
compared our key-value store design against Sherman~\cite{wang-sigmod-2022,
sherman-repo} and the MicroDB from Scythe~\cite{lu-taco-2024, scythe-repo},
two state-of-the-art RDMA key-value stores. We also compare against
Redis-cluster~\cite{redis} as a non-RDMA baseline. Results are shown in Figure
\ref{fig:kvstore-perf}. We measured throughput on read-only, mixed read-write,
and write-only operation distributions, across both uniform and Zipfian
($\theta=0.99$) key distributions, and across different node counts and per-node
thread counts. Each data point is the geometric mean of 5 runs with a 20
second duration, not including prefill.

All benchmarks use a 10MB keyspace, filled to 80\% capacity with 64-bit keys and
values. All benchmarks use the CityHash64 key hashing function~\cite{city-hash},
and the YCSB-C implementation of a Zipfian distribution~\cite{ycsb-c}.
We modified Sherman to issue a zero-length read fence between
lock-protected writes and lock releases to solve a bug related
to consistency issues (Section~\ref{ssec:consistency}). Our
\code{kvstore} also issues a fence for the same reason. For both, 
this fence incurs a 15\% overhead.

For LOCO, Sherman, and Redis, write operations are updates. For Scythe, we found
that stressing update operations led to program
instability and very low throughput, so we use the performance of insertion
operations as an upper bound on write performance.
For Redis, we configure a cluster with no replication or persistence. Since each
Redis server instance uses 4 threads, we create \code{ceil(num_threads/4)}
server instances for a given thread count. We use Memtier~\cite{memtier} as a
benchmark client. Each node runs
a single Memtier instance with threads equal to the thread count,
and 128 clients per thread (matching the LOCO large window size).

In addition, all systems expose a parameter we call the \emph{window
size}, which specifies the maximum number of outstanding operations per
application thread (note this is not a batch size -- each operation is
started and completed individually). 
Increasing LOCO's window size to 128 yielded
significant improvement (the "large window"
series). However, increasing Sherman's and Scythe's window sizes appeared to
cause internal errors, so the main results for all systems except Redis (see
above) use a window size of 3 for accurate comparison.

LOCO outperforms Sherman on read-only configurations. We believe this is
because Sherman reads whole sections of the tree from remote memory, while the
LOCO design looks up the location locally and only remotely reads the value.
On the other hand, LOCO's advantage over Sherman for Zipfian writes likely
comes from the better performance under contention of our ticket lock-based
design (Section \ref{ssec:more-channels}), compared to their
test-and-set-based design. 

Sherman outperforms LOCO (with a window size of 3) on mixed read-write and
write-only distributions on uniform keys, while the reverse is true for Zipfian
keys. Sherman's advantage here is likely due to the fact that, unlike LOCO,
Sherman colocates locks with data, allowing them to issue lock releases in a
batch with writes.

As a whole, these results demonstrate that LOCO can expose the full bandwidth of
RDMA to applications as a library just as effectively as ad-hoc, tightly-coupled
designs. 

\section{Conclusion}
\label{sec:conc}

In this paper, we describe LOCO, a library for building composable and reusable
objects in network memory. Our results show that LOCO can expose the full
performance of underlying network memory to applications, while easing
implementation burden on developers.  This work was partially funded
by an industry partner which provides hardware validation services using the
harness described in Appendix~\ref{app:powcon}.  One author also privately
contracted with this company to assist with commercialization efforts of this
harness.

\FloatBarrier
\appendix
\section{Backend}
\label{sec:backend}

Section~\ref{sec:design} described the channel system implemented by LOCO, and
Section~\ref{sec:channels} described a set of basic channels. In this section,
we briefly describe key features of the LOCO RDMA backend, which we have tuned
extensively to expose the full performance of RDMA to LOCO applications
(Section~\ref{sec:applications}).

The LOCO backend uses the \code{libibverbs} library for RDMA communication, and
the \code{librdmacm} library to manage RDMA connections. Both of these libraries
are components of the Linux \code{rdma-core} project~\cite{rdma-core}. LOCO
currently supports only RoCE~\cite{rdma-rocev2-2014} as a link layer, although
the only element missing for InfiniBand support is an implementation of the
connection procedure. The current design assumes a reliable, static network of
IP-addressable peers specified at application startup (the \code{hostnames} map
declared at line \ref{cd:host-map} of Figure \ref{lst:bar-lat}).

\subsection{Local Scalability}

LOCO implements multiple features aimed at increasing the scalability of
performing RDMA operations across multiple local threads. First, each thread in
a LOCO application uses a private set of QPs (one per peer), to avoid
unnecessary synchronization when multiple threads perform RDMA operations
simultaneously.  Second, all completions are delivered to a single completion
queue, which is monitored by a dedicated \textit{polling thread}, in order to
avoid contention on the completion queue.

Application code can monitor the progress of one or more operations by
registering an \code{ack_key} object with the polling thread, which provides
APIs for polling and waiting on completion of the operation. Internally, the
\code{ack_key} is a lock-free bitset with bits mapped to in-progress
operations. As operations complete, the polling thread clears the corresponding
bits, so that checking for completion of an \code{ack_key}'s registered
operations simply consists of testing whether the internal bitset is equal to
zero (i.e., empty). This approach avoids explicit synchronization between the
polling thread and application threads waiting for operations to complete.

\subsection{Network Memory Management}

Another important service the backend provides is management of each node's
network memory. Memory must be registered with \code{libibverbs} before it can
be accessed remotely. Since registration of a memory region incurs
non-negligible latency, we aggregate all registered memory used by LOCO channels
into a series of 1GB huge pages, each of which corresponds to a single
\code{libibverbs} memory region. The named memory region objects constructed by
channels each correspond to a contiguous sub-range of one of these regions.
Using huge pages reduces TLB utilization, which can have a significant
performance impact on multithreaded applications~\cite{chen-isca-1992}.

In addition to memory regions explicitly created by channels, we found it useful
to create a primitive for allocating temporary chunks of network memory used as
inputs and outputs of channel methods, which we call \code{mem_refs}.  We
allocate of backing memory for these objects from a per-thread pool of
fixed-size block, which are in turn allocated from the larger pool of registered
memory described above.

Finally, LOCO also provides the capacity to allocate local memory regions backed
by \emph{device memory}, which resides on the network card.  RDMA accesses to
device memory are faster than those to system memory, since they are not
required to traverse the PCIe bus to main memory. However, since device memory
is not coherent with main memory, it is mainly useful for holding state
exclusively accessed through the network, such as mutex state.

\section{Distributed DC/DC Converter System}
\label{app:powcon}

As an additional application of LOCO, we implemented a
model of a hardware control loop which exploits its low latency.

\subsection{System Design}

\begin{figure}
\includegraphics[width=0.45\textwidth]{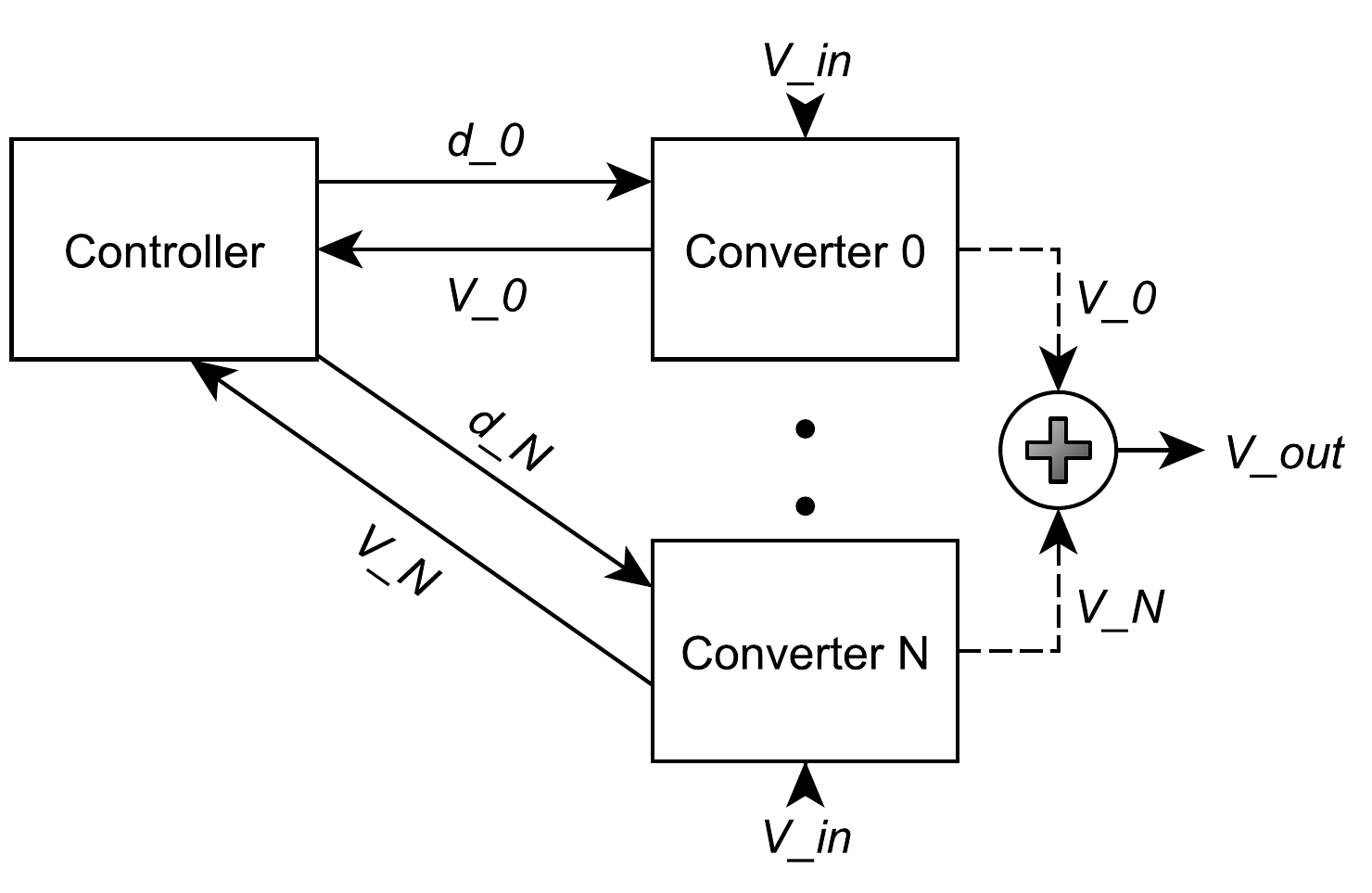}

\caption{Schematic of the system modeled by the \code{power_controller} channel.
Solid arrows represent LOCO \code{owned_vars}, and dashed arrows represent
electrical connections. \code{d_N} represents the duty cycle parameter used to
control converter \code{N}, and \code{V_N} represents the output voltage at
converter \code{N}.}

\label{fig:powcon}
\end{figure}

An additional application channel we have implemented is the
\code{power_controller}, a real-time simulation of a distributed DC/DC converter
system controlled by a discrete-time control
loop~\cite{corradini-textbook-2015}. The simulation (Figure \ref{fig:powcon})
consists of a single machine which acts as a \emph{controller}, and an arbitrary
number of machines simulating the physical characteristics of a
\emph{converter}. The role of the controller is to regulate the duty cycles
($d$) of the converters, which are supplied with a steady input DC voltage, to
produce a target output voltage ($V_{ref}$). The converters return voltage
values ($V$) which are used to calculate the next setting of their duty cycles,
closing the control loop.

The \code{power_controller} channel consists of two arrays of \code{owned_vars}
representing the duty cycle (owned by the controller) and output voltage (owned
by the converter) for each converter. The participating machines run fixed-time
loops: each loop iteration at a converter calculates a new simulated $V$ and
pushes it to the controller, while each iteration at the controller calculates a
new $d$ for all controllers based on their most recent $d$ and $V$ values. The
overall output voltage of the system at each step (as seen by the controller) is
the sum of all converters' most recent output voltage.

Network memory is a good fit for this application because it is highly sensitive
to network latency; with the parameters we have chosen, the output will only
converge if latency of the control and feedback messages is consistently less
than 40 $\mu$s. This requirement would be difficult to meet with traditional
message-passing protocols: while a protocol such as UDP can easily achieve this
latency on an uncontended network, it would be difficult to manage the
scheduling jitter, copying, and cache contention in the software network
protocol stack. 

An extension of this control loop harness was developed in LOCO for validating
hardware components such as the power controller and converters within partially
simulated environments (hardware-in-loop testing). The system is currently in
beta testing for production use, with expected commercial release later this
year. 

\subsection{Evaluation}

\begin{figure}
\includegraphics[width=.45\textwidth]{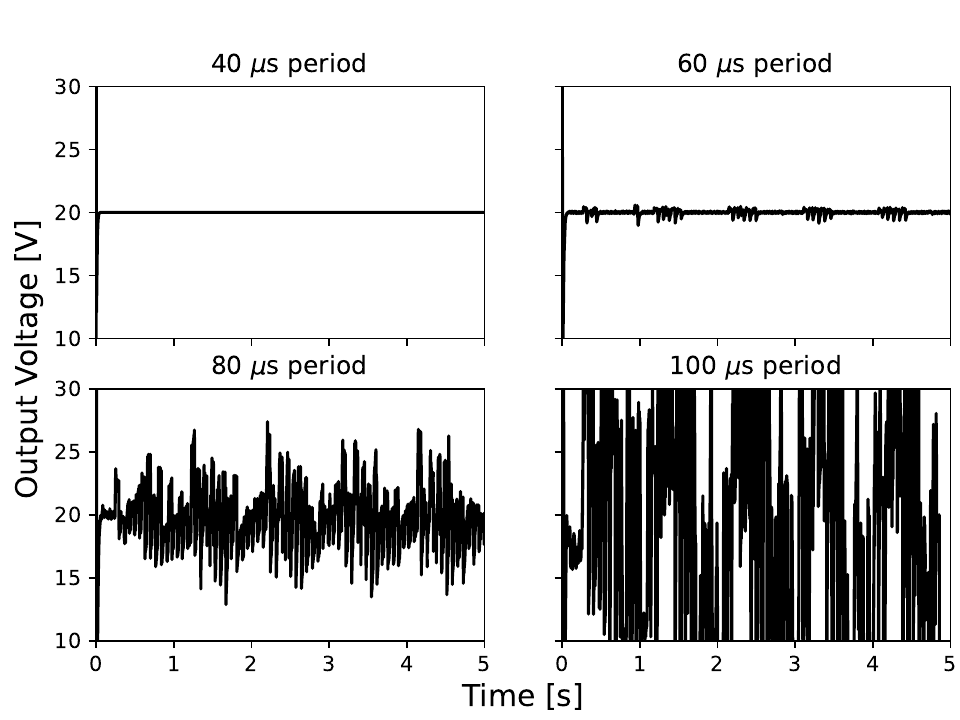}

\caption{Output voltage for the DC/DC converter simulation at various control
loop frequencies. 
}

\label{fig:powcon-perf}
\end{figure}

To evaluate whether LOCO meets the latency requirements of this system, we
instantiated a cluster with one controller and 20 converters and measured the
output voltage over time at various loop periods. The effect of changing the
loop period is to simulate higher link latency, since we cannot increase the
latency of the RDMA link. The loop period at the converters is fixed at 10
$\mu$s to approximate the continuous nature of their transfer function. We ran
each simulation for 5 seconds.

The system parameters are selected to maintain a stable output voltage with a
controller loop period of 40 $\mu$s or lower. The increasing instability in the
output resulting from increasing the loop period past this value is clearly
visible in Figure \ref{fig:powcon-perf}. The series with period greater than 40
$\mu$s also exhibit large transients at simulation start. These are
mostly invisible on the plots due to their brief duration, but would be
unacceptable in a real system.

\section{Safety}
\label{sec:proof}

The \pcode{kvstore} object described in section~\ref{sec:applications} is
linearizable~\cite{herlihy-toplas-1990}, and we here provide an informal proof
of safety.  Note that our proof leverages both the composition of
linearizability and the mutual exclusion property of our locks, their use are
simplified by the composable nature of LOCO channels.

\subsection{Preliminaries}

We choose linearization points~\cite{herlihy-toplas-1990} 
for each modification operation as follows.
A \code{write} linearizes when the key, value, and checksum
are fully placed on the host node.  A \code{delete} linearizes
when the \code{valid} bit is unset (before all nodes have
modified their local index and acknowledged the deletion).
An \code{insert} linearizes when the \code{valid} bit is set
(after all nodes have
modified their local index and acknowledged the insertion).

The linearization points of reads are determined retrospectively
depending on the read value.

Investigation of the algorithm determines every read consists of two steps (possibly repeated).
(1) A fetch from the local index to determine the node and address of the key's associated value.
(2) A remote read to this location.
The remote read can result in one of three possible scenarios.
\begin{enumerate}
\item If the read contents match the associated counter and checksum and the valid bit is set, the read linearizes
at the point of the remote read's execution and returns the read value.  
\item If the read contents and the associated checksum do not match, 
the read overlaps with an ongoing (torn) update, and the read is retried in its entirety.
\item If the read contents match the requested counter but the valid bit is unset, this implies
either that an in-progress insert has not yet linearized, or an in-progress
deletion has already linearized but not yet updated the local index. The read linearizes
at the point of the remote read's execution and returns \code{EMPTY}.
\item If the read contents do not match the requested counter, this implies an in-progress
\code{delete} has completed but had not yet updated the local index when the read was initiated,
and later operations have reused the slot.  In this case, the
read linearizes immediately after the delete and returns \code{EMPTY}.
\end{enumerate}

\subsection{Proof of Safety}

\begin{lemma}
\label{lm:tmo}
All \code{write}s, \code{delete}s, and \code{insert}s
for a given key form a total modification order which respects the real-time
ordering of the operations.
\end{lemma}

\begin{proof}
By mutual exclusion on the per-key lock, each operation's
effects are completed before any subsequent operation.
\end{proof}

\begin{lemma}
\label{lm:read}
Every \code{read} returns a value consistent with the total modification order
and which respects real-time ordering of the operations.
\end{lemma}

\begin{proof}
We break our proof into three cases contingent on the
result of the remote read.  In the first, the local index counter
matches the result of the remote read, in the second, the local
index does not match, in the third, the checksum does not match
and the read cannot determine the case.  We validate the
linearization of the read for each case in reverse order.

In the case where the checksum does not match, this is an
atomicity violation, and the operation retries without linearizing.

In the case where the local index does not match,
the counter value read by the remote read indicates that the local
index is out of date.  This case 
implies an in-progress
\code{delete} has linearized but not yet updated the local index, 
and later operations have reused the slot.  As the remote \code{delete}
cannot complete until the local index is updated, the read must have
overlapped in real-time with the delete, and thus can return \code{EMPTY}.

In the case where the local index matches, the remote read may discover
a either a valid or invalid value.  If the value is valid,
the read can return the read value, as this value respects
the most recent linearization of a modification to the location.
If the read discovers an invalid flag --- this indicates
that its local index is out-of-date with respect to an ongoing
\code{delete} or \code{insert}.  Returning \code{EMPTY} respects
the linearization point of both operations (note the asymmetry of the
modifying operations to enable this possibility).
\end{proof}

By lemma~\ref{lm:tmo} and~\ref{lm:read}, and by composition of linearizable
objects~\cite{herlihy-toplas-1990},
\begin{theorem}
The presented hashmap is linearizable.
\end{theorem}

\bibliographystyle{plain}
\interlinepenalty=10000
\bibliography{paper}

\end{document}